\documentclass[conference]{IEEEtran}
\IEEEoverridecommandlockouts
\usepackage{cite}
\usepackage{amsmath,amssymb,amsfonts}
\usepackage{algorithmic}
\usepackage{graphicx}
\usepackage{textcomp}
\usepackage{xcolor}
\def\BibTeX{{\rm B\kern-.05em{\sc i\kern-.025em b}\kern-.08em
    T\kern-.1667em\lower.7ex\hbox{E}\kern-.125emX}}
\begin{document}

\title{Passivity based Stability Assessment for Four types of Droops for DC Microgrids  \\

\thanks{This work was supported in part by the Advanced Research Projects Agency-Energy (ARPA-E), U.S. Department of Energy, under Award DE-AR0001580.}
}

\author{
  \IEEEauthorblockN{Muhammad Anees}
  \IEEEauthorblockA{\textit{FREEDM System Center} \\
    \textit{North Carolina State University}\\
    Raleigh, USA \\
    manees@ncsu.edu}
\and
        \IEEEauthorblockN{Lisa Qi}
  \IEEEauthorblockA{\textit{ABB Corporate Research Center} \\
    \textit{ABB Inc.}\\
    Raleigh, USA \\
    lisa.qi@us.abb.com}\\

  \IEEEauthorblockN{Srdjan Lukic}
  \IEEEauthorblockA{\textit{FREEDM System Center} \\
    \textit{North Carolina State University}\\
    Raleigh, USA \\
    smlukic@ncsu.edu}   
\and  
   \IEEEauthorblockN{Mario Schweizer}
  \IEEEauthorblockA{\textit{ABB Corporate Research Center} \\
    \textit{ABB Inc.}\\
    Baden-Daettwil, Switzerland\\
    mario.schweizer@ch.abb.com}
}

\maketitle

\begin{abstract}
DC microgrids are getting more and more applications due to simple converters, only voltage control and higher efficiencies compared to conventional AC grids. Droop control is a well know decentralized control strategy for power sharing among converter interfaced sources and loads in a DC microgrid. This work compares the stability assessment and control of four types of droops for boost converters using the concept of passivity. EN standard 50388-2 for railway systems provides a reference to ensure system stability in perspectives of converters and system integration. Low pass filter (LPF) in the feedback of the droop control is used to ensure converter passivity. Bus impedance is derived to ensure system passivity with less conservativeness. Analytical approach for design of passive controller for all four types of droops is verified through time domain simulations of a single boost converter based microgrid feeding a Constant Power Load (CPL).
\end{abstract}
\begin{IEEEkeywords}
DC Microgrids, Droop, Stability, Passivity
\end{IEEEkeywords}

\section{Introduction and Literature Review}

Different types of microgrids are based on the nature of electrical currents; AC, DC, or Hybrid microgrids. DC microgrid is considered simple and less complex than AC microgrids from a control perspective. Moreover, the inherent DC nature of renewables (PV and Wind etc.) and energy storage technologies (Batteries and fuel cells) makes DC microgrid a preferable choice from an integration perspective [1].
DC Microgrids are gaining more and more fame due to enhanced availability of the DC resources including energy generation (Photovoltaics and wind), storage (supercapacitors and batteries) and consumption (DC loads like HVAC, Lighting, DC fans etc.). Due to resources availability and improved efficiency in DC systems, DC microgrids have applications in diverse settings including data centers, off grid electrifications, telecommunications, electric vehicle charging stations, controlled indoor farming, maritime and offshore applications, military installations, commercial and residential buildings.
\subsection{Hierarichal Control of DC Microgrids}

Interconnection of the resources and loads in microgrids is possible through power electronics converters; regulation of the DC bus voltages through these converters is a primary control demand. The hierarchal control approach ensures the stable operation of microgrids, where each control layer is responsible for specific responsibilities based on the dynamic nature of responsibilities [2]. Hierarchal control for DC microgrids is given in Fig. 1. In Fig. 1, on the left, from bottom to top, are the control layers from device level to system level with objectives mentioned in the triangle. While on the right, from bottom to top, is the bandwidth of each control loop. Bandwidth of the lowest control loop is usually in microseconds – as a rule of thumb (6-10) times slower than the converter's switching frequency. Examples of tracking control for converters are current control, voltage control and dual loop control [2]. As a rule of thumb, each outer control layer of the converter control is kept (6-10) times slower than the inner control layer. Following this hierarchy, outer voltage loop will be (6-10) times slower than the inner current loop and droop loop will be (6-10) times slower than outer voltage/current loop. 
\begin{figure}[htbp]
\centerline{\includegraphics{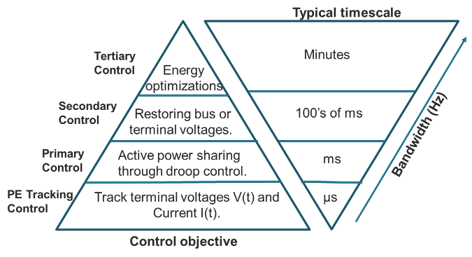}}
\caption{Hierarichal Control of DC Microgrids}
\label{fig}
\end{figure}
\subsection{Droop Control}
The idea of the droop control came from the AC generator, where the output power transfer from the generator is modulated with the frequency of operation. Following the same approach from the AC generator, droop control in converters of DC microgrids uses the terminal voltages of the converter as a signal for the intended power flow throughout the microgrid. There are multiple implementations of the droop reported in the literature [2-10]. From feedback variables perspective, droop can be categorized as power-based droop (VP and PV droop with droop coefficient k) and current based droop (IV and VI droop with droop coefficient d) [3]. There exists a fundamental conflict of uniform power sharing with voltage regulation because of non-uniform line impedances [5]. To resolve this conflict and improve the overall performance of the droop, different researchers proposed different methodologies [4-15]. Some of the approaches required communication and some don’t. Communication enables the controllers for a coordinated action to regulate an average voltage and average power sharing correspondingly [5-7]. Decentralized approaches include the use of non-linear or adaptive droop gains [8-10]. Communication also enables the primary/droop controller to coordinate with the secondary controller using hierarchical control, which can use centralized or distributed control approach for the uniform power sharing along with the restoration/regulation of bus voltages [10-13]. 

\subsection{Passivity based Droop Control Design}
From a converter inner current control perspective, inductor current is controlled; in case of boost converter, inductor current is not same as output current. Therefore, for boost converter, both the inductor (input) and output current are measured. For a typical boost converter, all four types of droops can be represented by the following equations. 
\begin{equation}
    VI - Droop:      { v=v_o-d*i} 
\end{equation}
\begin{equation}
   VP - Droop:     { v=v_o-k*P} 
\end{equation}
\begin{equation}
   IV - Droop:      i_0^* = \frac{v_o-v} {d} ,  i_L^* = { (i_0^*-i_o)*G_{cc}(s) } 
\end{equation}
\begin{equation}
    PV - Droop:      i_0^* = \frac{v_o-v} {k*v} ,  i_L^* = { (i_0^*-i_o)*G_{cc}(s) } 
\end{equation}

Voltage \textit{v }in case of VI and VP droop will be fed to outer voltage loop of standard dual loop control. While reference current  in case of IV and PV, will be fed to outer current loop that translate the output current to inductor current with \textit{Gcc(s)} as current controller in case of boost converter.

Knowing the structure of control loops, a small signal equivalent output impedance is deduced for four types of droops (IV, VI, PV, VP) for boost converter – sets of equations in Appendix I. The deduced equivalent small signal impedance has been verified through the time domain frequency sweeping simulations for each type of droop. Detailed derivation and verification through time domain simulation is not added to the paper due to pages limit but can be provided upon request.

Each of the above-mentioned droop type has their own advantages or disadvantages, but a detailed stability performance comparison of all four droops is still missing. At the same time passivity-based control design is currently a hot topic of research to ensure microgrid stability. Many papers have discussed DC system stability using passivity theorem either from converter design or system design perspectives [15-19]. Major contributions of this paper include 1) guidance for converter manufacturers and system integrators combining the European standard EN50388-2 and bus impedance passivity requirements [14] for stable DC microgrids, and 2) a comparison of performance (alternatively passivity characteristics) for four types of droops under varying microgrid conditions/parameters. Another important contribution is that Low pass filter bandwidth in primary droop control is used to make sure the both converter and system are passive, while the hierarichal control design remains intact. 
\section{Converter and System Passivity}
Following the European standard EN50388-2 for AC traction system, the responsibility to maintain system stability is shared between the converter manufacturers and the infrastructure operator in an organizational tractable manner by the compliances with two rules [14]. Similarly, in DC systems, it is desirable to design the converter control and the overall system integration in such a way that we are ensuring the passivity of the converter and system. 
\begin{itemize}
\item Equivalent converter impedances are strictly passive above a frequency threshold (87Hz in 16.6Hz railway; 300Hz in 50Hz applications).

\textit{\textbf{“strictly passive” }means that the complex converter impedance has positive real part for all frequencies above the threshold}
\item The grid is not allowed to have weakly-damped passive resonances below the frequency threshold, because converters are allowed to be non-passive there.
\end{itemize}

The above two requirements from EN50388-2 together guarantee system stability and can be used as converter and system design guideline without knowledge of the complete system, because parallel connected systems preserve passivity in the respective frequency range. Also, when more and more passive converters are added into the system which are passive with certain margin, the aggregated behavior of all converters is still passive. The concept to ensure converter passivity and system passivity for AC traction system stability can also be applied to DC microgrid stability. The frequency threshold of 300Hz may not be true in DC microgrid and can be derived from rigorous analysis.
	
The two requirements ensure system stability by 1) passive converters provided by converter manufacturers and 2) not allowing weakly-damped passive components in system integration by system integrators. The weakly-damped passive components in system integration is a conservative way to ensure system stability. Instead, bus impedance $Z_{\text{bus}}$ offer an integrated way to ensure system passivity using the combination of all parallel impedances of the connected converters to that bus [15]. For m+n converters (m load converters , n source converters); equivalent bus impedance $Z_{\text{bus}}$ can be computed as 
\begin{equation}
Z_{\text{bus}} = Z_1 \parallel (Z_2 \parallel \dots \parallel Z_n) \parallel \dots \parallel (Z_{n+1} \parallel \dots \parallel Z_{m+n})
\end{equation}

For system stability, the equivalent $Z_{\text{bus}}$ should follow the passivity criteria, which can be translated as: Th\textit{e phase angle of $Z_{\text{bus}}$ is between $\pm 90^\circ$ for the entire range to ensure passivity.}

\begin{table}[htbp]
\caption{System Parameters}
\begin{center}
\begin{tabular}{|c|c|}
\hline
\textbf{\textit{Parameters}} & \textbf{\textit{Value}} \\
\hline
Global no-load voltage ref [Vo] & 350V \\
\hline
Source Voltages [E] & 130V \\
\hline
Load Power & 3600W \\
\hline
Source resistance [rBAT] & 0.03$\Omega$ \\
\hline
Inductance [L] & 2mH \\
\hline
Inductor ESR [r] & 0.01$\Omega$ \\
\hline
Per meter line resistance [R] & 10m$\Omega$ \\
\hline
Per meter line inductance [L] & 10uH \\
\hline
Line length [LL] & 29m \\
\hline
Bus Capacitance [C] & 3.3mF \\
\hline
VI/IV droop coefficient [d] & 1 \\
\hline
VP/PV Droop coefficient [k] & 10V/3600W \\
\hline
Current control bandwidth & 3KHz \\
\hline
Voltage control bandwidth & 200Hz \\
\hline
Outer current control bandwidth & 200Hz \\
\hline
Switching frequency [fsw] & 20KHz \\
\hline
\end{tabular}
\label{tab1}
\end{center}
\end{table}

Different researchers used different approaches for passivity-based converter controller design for DC microgrid [15-19]. Etc. Herein this paper, a conventionally used LPF implemented in the feedback of the droop control is used to make sure the converter is passive. There are different advantages of using LPF as a tweaking parameter to ensure system passivity. The most important one is that the tracking control (design of inner and outer loops) remains intact, while in other cases we need additional loops or filters that are directly impacting the other loops. Another important advantage is the emulation of virtual inertia in DC system through LPF BW selection – equivalent machine behavior using droop with LPF [20]. 
\begin{figure}[htbp]
\centerline{\includegraphics[scale=0.9]{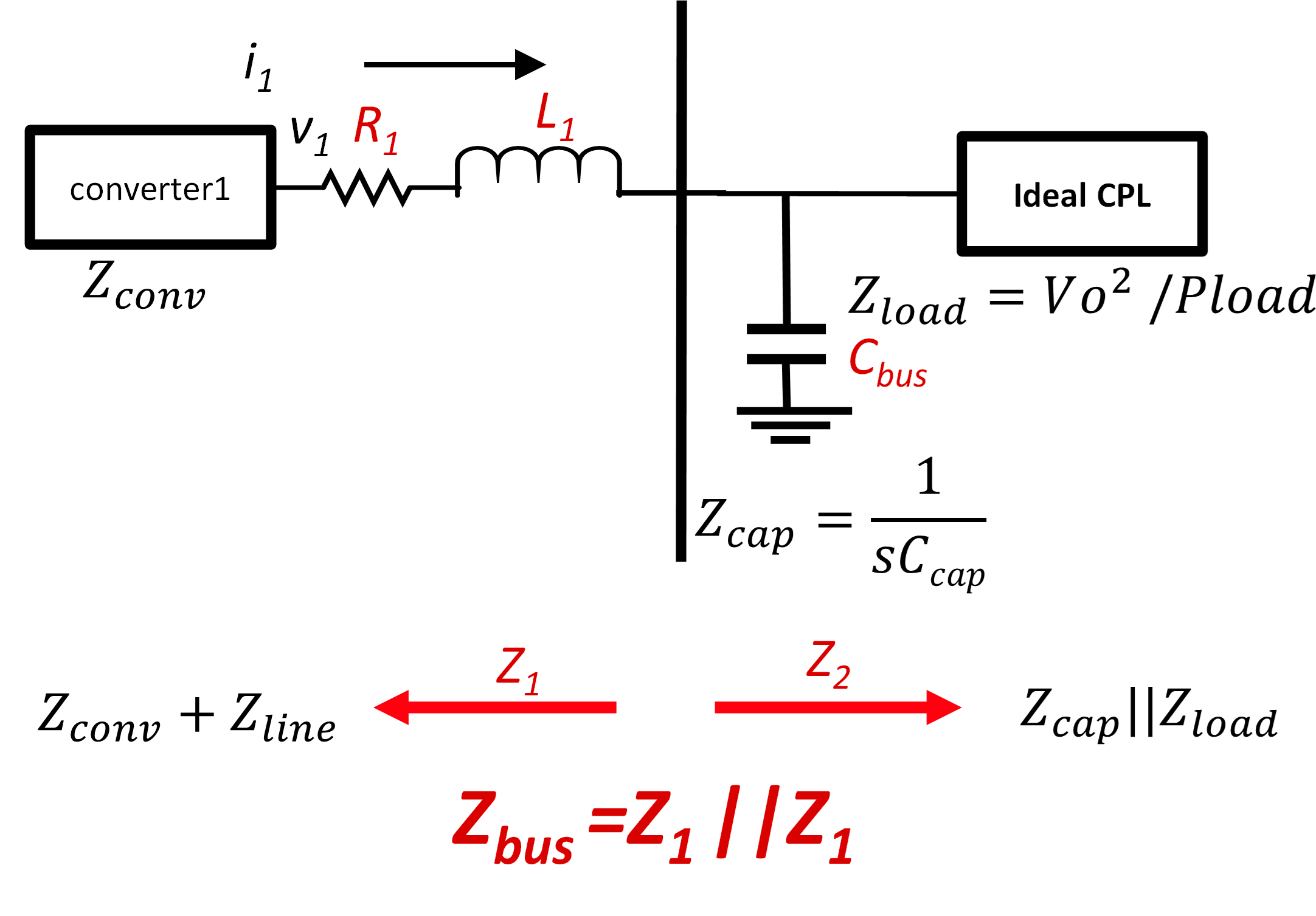}}
\caption{Microgrid Model with CPL}
\label{fig}
\end{figure}
Using a simple microgrid of Fig. 2 as the test system, a boost converter powers a CPL through line impedance. Converter and microgrid passivity will be assessed using the specified parameters of Table 1. The converter's impedance plot displays non-passive regions around the current control loop bandwidth if no filter or LPF cutoff frequency is equal or above 50Hz despite classical control loop designs. Careful LPF bandwidth selection can ensure converter passivity. At the load input, a bus capacitance is present along with CPL. System analysis shows that even the converter becomes non-passive beyond a 50Hz bandwidth (Fig. 3), while the overall microgrid bus impedance remains passive. With bus impedance, even though there are some slight violations of passivity for converters, the system stability can still be ensured. This fact highlights the passivity of bus impedance to provide system stability while allowing some margins for converter design (Fig. 4).

\begin{figure}[htbp]
\centerline{\includegraphics[scale=0.85]{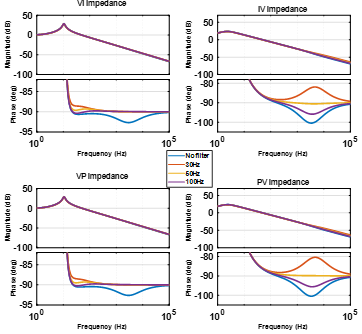}}
\caption{Converter impedance around current loop BW}
\label{fig}
\end{figure}
\begin{figure}[htbp]
\centerline{\includegraphics[scale=0.9]{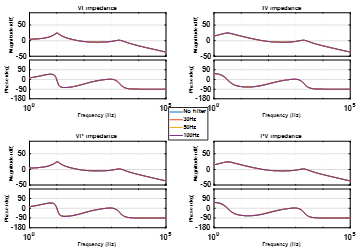}}
\caption{Microgrid or bus impedance}
\label{fig}
\end{figure}
\begin{figure*}[htbp]
\centering
\includegraphics[scale=1]{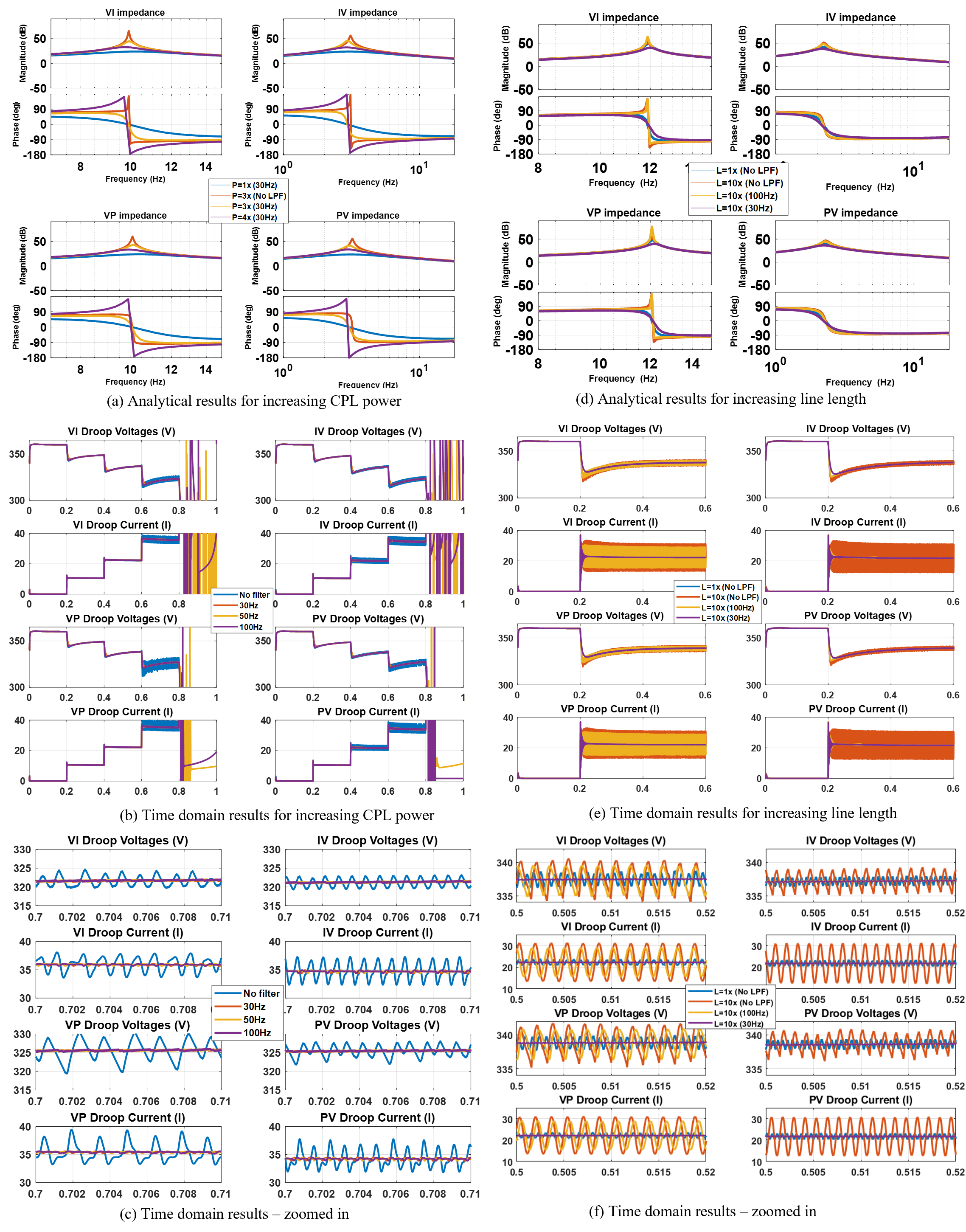}
\caption{Effect of CPL power and line impedance on microgrid passivity}
\label{fig}
\end{figure*}

\section{Parameters Impacting Microgrid Passivity}
There are different parameters which can impact the bus impedance, including the CPL power level, line impedance and bus capacitance. Now we will play with CPL power and line impedance to see its impact on bus impedance passivity. The bus capacitance case is ignored for this paper due to space constraint. Analytical findings are supported by time domain simulations, utilizing system parameters, unless otherwise noted.

\subsection{Impact of CPL Power on System Passivity}
Increasing the CPL power pushes the bus impedance to non-passive region, as observed in Fig. 5(a). At the standard rated power (P=1x), all four droops exhibit complete passivity with a 30Hz filter. However, when the CPL power increases to 3x (without LPF), all four droops show non-passive regions, necessitating at least a 30Hz LPF to eliminate the non-passive behavior. But the same 30Hz LPF doesn't suffice when the CPL power further escalates to 4x. In Fig. 5(b), time domain simulations reveal instability for all four droops at P=4x, with the 30Hz filter stabilizing oscillations at P=3x, as shown in the zoomed-in Fig. 5(c). These results delineate the CPL power's limit to destabilize the system from stable setpoints. Furthermore, the time domain simulations illustrate minor oscillations at P=2x for IV and PV, but not for VI and VP. Consequently, for higher powers, VI and VP appear to be preferable choices compared to IV and PV.
\subsection{Impact of Line Impedance on System Passivity}
This case studies how line impedance impacts system passivity and the role of LPF in managing oscillations due to high inductive line impedance. With the line length increased to 10x and line inductance per meter set to 10uH 
at a P=2x system operation level, Fig. 5(d) shows non-passive areas in the bus impedance for all four droop types at this increased line length. The corresponding time domain results in Fig. 5(e) reveal that the higher line impedance (without LPF) causes oscillations, aligning with non-passive zones from the analysis. A 100Hz filter controls these oscillations for IV and PV droops, while a 30Hz filter is needed for VI and VP. Zooming in (Fig. 5(f)) confirms that a 100Hz filter ensures full system passivity for IV and PV droops, not for VI and VP. For longer or more inductive lines, IV and PV droops are more suitable than VI and VP. The proposed approach in the article is verified on a single bus microgrid. The authors believe that this approach can be extended to multi-node and more complex DC microgrids as part of future research.

\section{Conclusion}
This article provides a baseline approach for the assessment of system stability using passivity for droop-controlled DC microgrids. First, the passivity of the converter and microgrid is explained in the context of the EN standard 50388-2 for traction systems, the concept of bus impedance, and state of the art literature. Then a detailed approach for the assessment of the passivity of a converter/microgrid is provided. Conventional LPF is used to ensure the converter and microgrid passivity, which have major advantage of virtual inertia emulation and inner tracking control remains intact. Then a tweak for microgrid parameters is performed for four types of droops and it is found that for high power levels VI and VP is a preferable choice, IV and PV is a preferred choice for high line lengths and inductances. 

\vspace{2cm} 

\begin{appendices}

\section*{Appendix I}
\begin{center}
Notation followed from Table 1, where, $Z(s)= r_{\text{BAT}} +r_f+Ls$ , $d=R_{\text{vd}}$ , $k_{pv}= k_{vp} = k$ , $Dc= \text{duty cycle}$
\end{center}

\begin{equation*}
\scalebox{0.99}{$
Z^{\text{PV}} = \frac{\left( (1-Dc) + I_L G_c (s) \right) V_{\text{BUS}} G_c (s) G_{cc} (s) + \left( 1-I_L G_c (s) G_{cc} (s) \right)(Z(s)+V_{\text{BUS}} G_c (s))}{\left[(1-Dc)+I_L G_c (s)\right]\left[(1-Dc)+\left(V_{\text{BUS}} G_c (s) G_{cc} (s) V_{\text{ref}}\right)/(k_{pv} V_{\text{BUS}}^2)\right] + \left[Z(s)+V_{\text{BUS}} G_c (s)\right]\left(sC -(I_L G_c (s) G_{cc} (s) V_{\text{ref}})/(k_{pv} V_{\text{BUS}}^2)\right)}
$}
\end{equation*}

\begin{equation*}
Z^{\text{IV}} = \frac{\left( (1-Dc) + I_L G_c (s) \right)V_{\text{BUS}} G_c (s) G_{cc} (s) + \left( 1-I_L G_c (s) \right)(Z(s)+V_{\text{BUS}} G_c (s))}{\left[(1-Dc)+I_L G_c (s)\right]\left[(1-Dc)+(V_{\text{BUS}} G_c (s) G_{cc} (s))/R_{\text{vd}}\right] + \left[Z(s)+V_{\text{BUS}} G_c (s)\right]\left(sC-(I_L G_c (s) G_{cc} (s))/R_{\text{vd}}\right)}
\end{equation*}

\begin{equation*}
Z^{\text{VI}} = \frac{[(1-Dc) V_{\text{BUS}}-I_L Z(s)]R_{\text{vd}} G_c (s) G_v (s)+V_{\text{BUS}} G_c (s)+Z(s)}{[(1-Dc) V_{\text{BUS}}-I_L Z(s)] G_c (s) G_v (s)+[V_{\text{BUS}} G_c (s)+Z(s)]sC+(1-Dc) I_L G_c (s)+(1-Dc)^2 }
\end{equation*}

\begin{equation*}
Z^{\text{VP}} = \frac{[(1-Dc) k_{vp} V_{\text{BUS}}^2-I_L Z(s) k_{vp} V_{\text{BUS}}] G_c (s) G_v (s)+V_{\text{BUS}} G_c (s)+Z(s)}{[(1-Dc) V_{\text{ref}}-V_{\text{ref}}/V_{\text{BUS}}  I_L Z(s)] G_c (s) G_v (s)+[sCV_{\text{BUS}}+(1-Dc) I_L ] G_c (s)+sCZ(s)+(1-Dc)^2 }
\end{equation*}

\end{appendices}

\end{document}